\documentclass[%
 superscriptaddress,
 reprint,
 amsmath, amssymb,
 prb,
 citeautoscript,
 floatfix
]{revtex4-1}
\usepackage{graphicx}
\usepackage{placeins}
\usepackage{dcolumn}
\usepackage{bm}
\usepackage{tikz}
\usetikzlibrary{shapes.geometric, arrows.meta, positioning, calc, fit, backgrounds}
\usepackage[utf8]{inputenc}
\usepackage{xcolor}
\usepackage{soul}
\usepackage{algorithm}
\usepackage{algpseudocode}
\usepackage{booktabs}
\usepackage{overpic}
\usepackage{xr}
\algrenewcommand\algorithmicrequire{\textbf{Input:}}
\algrenewcommand\algorithmicensure{\textbf{Output:}}
\algnewcommand{\LineComment}[1]{\State \(\triangleright\) \textit{#1}}

\newcommand{\vG}{\mathbf{G}}
\newcommand{\vR}{\mathbf{R}}
\newcommand{\bvr}{\mathbf{r}}
\newcommand{\vb}{\mathbf{b}}
\newcommand{\vc}{\mathbf{c}}
\newcommand{\vq}{\mathbf{q}}
\newcommand{\mA}{\mathbf{A}}

\begin{document}
\preprint{APS/123-QED}
\title{\textit{opt}-DDAP: Optimisable density-derived atomic point charges\\via automatic differentiation}
\author{Mohith H.}
\author{Sudarshan Vijay}
\affiliation{Department of Chemical Engineering, Indian Institute of Technology Bombay, Powai, Mumbai, Maharashtra 400076, India}
\email{sudarshan.vijay@iitb.ac.in}
\date{\today}

\begin{abstract}
Interatomic potentials which accurately describe long-range electrostatics require atom-centred charges.
One such method to determine these atom-centred charges from density functional theory (DFT) calculations is the density-derived atomic point (DDAP) charge method.
DDAP fits atom-centred Gaussians to the ground-state DFT charge density and preserves the multipole moments that govern long-range electrostatics.
While these charges accurately predict long-range behaviour, in practice, they are limited by their reliance on fixed, heuristic parameters and a constrained solver that becomes numerically unstable for complex or covalent systems.
In this work, we present \textit{opt}-DDAP, which solves this limitation by reformulating the algorithm as a differentiable computational graph.
This reformulation allows for the optimisation of Gaussian basis parameters and the reciprocal-space cutoff using automatic differentiation.
To ensure numerical robustness through this automatic differentiation process, we replace the conventional Lagrange-multiplier approach with a pseudo-inverse solution followed by charge renormalisation, maintaining stability even in the presence of ill-conditioned matrices.
We validate the framework on NaCl vacancy supercells and on MoS$_2$, demonstrating faithful reconstruction of both absolute and difference charge densities.
The optimised charges are intended to serve as inputs to effective electrostatic models in machine-learning and empirical interatomic potentials that incorporate long-range interactions.
\end{abstract}

\maketitle

\section{Introduction}\label{sec:intro}

Plane-wave density functional theory (DFT) calculations are the workhorse of computational materials science~\cite{Teale2022,vasp}.
High-throughput DFT calculations, based on automated workflows, have been used to generate large scale databases of outputs of DFT calculations consisting of quantities such as total energy, atom-centred forces and band structures \cite{materialsproject2013, curtarolo2012aflow}.
These databases have been used to predict novel battery materials and heterogeneous catalysts, develop mechanistic insight and create new design principles.\cite{mueller2016high, aykol2016network, winther2019catalysis, kitchin2018machine}

An output from DFT calculations that has received less attention are atom-centered charges.
These charges are post-processed from the ground-state DFT charge density.
Physically meaningful atom-resolved charges are essential for developing chemical insights of charge transfer reactions and parametrising accurate empirical force-fields.
More recently, these atom-centred charges have also been used to incorporate long-range electrostatic interactions in machine-learning interatomic potentials (MLIPs)~\cite{kocer2022}.

There are two key challenges to creating automated workflows to compute atom-centered charges.
First, the concept of an atom-centred charge from a ground-state charge density is inherently ambiguous.
Different charge partitioning schemes, such as the Bader charge scheme, Mulliken charges, natural bond orbital charges~\cite{bader1990, mulliken1955, glendening2012} and schemes based on projections rely on heuristics to compress the charge density into a point charge representation.
Second, the outputs of all schemes are inherently sensitive to the parameters chosen to split the charge density into an atom-centred charge. 

An alternative, practical approach to compute the atom-centred charges is through the density-derived atomic point charge (DDAP) method~\cite{blochl1995}.
Instead of partitioning the charge density into atom-centred regions, this approach constructs a superposition of atom-centred Gaussians, weighted so as to preferentially reproduce the low-$\vG$ components that encode the electrostatic multipole moments. By prioritising the monopole, dipole, and quadrupole terms in the fit, the resulting 
charges correctly capture the long-range electrostatic interactions that dominate at 
large separations, since the interaction energy between well-separated charge 
distributions is determined entirely by their multipole moments~\cite{blochl1995}. 
This minimal representation reproduces the self-consistently 
determined electrostatic potential, making it suitable for Ewald summations performed in molecular dynamics simulations, with applications ranging from parametrizing force fields for two-dimensional materials~\cite{govindrajan2018hbn, kumar2024bnnt} to
extracting bulk ionic charges in ionic liquids~\cite{schmidt2010ionic, kocer2022, dommert2012charge}. 

While DDAP charges solve the challenge of charge partitioning schemes, they are still sensitive to the parameters and numerical convergence.
Standard implementations solve a constrained linear system via Lagrange multipliers to enforce charge conservation.
This system computes the inverse of a Gaussian overlap matrix ($\mA$) whose condition number~$\kappa(\mA)$ grows rapidly with the number of Gaussians per atom centre (GPA) or with unfavourable choices of Gaussian widths, leading to catastrophically unphysical charges when
$\kappa(\mA) \gtrsim 10^{8}$.
Furthermore, the Gaussian basis parameters, i.e., number of Gaussians per atom, their decay lengths~$\sigma$, and the reciprocal-space cutoff~$g_c$, are typically chosen heuristically and held fixed.
Default values may not transfer across chemically distinct systems (ionic vs.\ covalent), different supercell sizes, or defected structures.

In this work, we solve this challenge of parameter sensitivity by developing an optimisable DDAP framework, \textit{opt}-DDAP.
Our method treats the Gaussian basis parameters as differentiable variables which are optimised by minimising the difference between the ground state and model charge density.
We achieve numerical robustness by replacing the constrained minimisation (through the Lagrange constraints) with an unconstrained minimisation by computing the Moore--Penrose pseudoinverse followed by charge renormalisation.
This manuscript is organised as follows: in Section \ref{sec:challenges}, we describe the challenges with picking parameters for the DDAP framework.
In Section \ref{sec:autoddap_algo}, we describe modifications to the DDAP framework in order to make it optimisable by auto-differentiation.
In Section \ref{sec:validation}, we validate the approach on NaCl bulk and vacancy systems and on MoS$_2$, and demonstrate that the optimised charges faithfully reconstruct both absolute and difference charge densities.

\section{Challenges with applying DDAP in practice}\label{sec:challenges}

\subsection{The DDAP formalism}\label{sec:formalism}

We start by describing the fitting parameters introduced in the original DDAP formulation~\cite{blochl1995}.
The self-consistent charge density of a periodic system is expressed as a Fourier series
\begin{equation}\label{eq:rhoG}
  \rho(\bvr) = \sum_{\vG} \rho(\vG)\, e^{i\vG\cdot\bvr},
\end{equation}
where $\vG$ are reciprocal-lattice vectors, $\rho(\mathbf{r})$  and $\rho(\vG)$ are the charge density in real and reciprocal space respectively.
DDAP attempts to capture $\rho(\mathbf{r})$ through a model density, $\hat{\rho}(\bvr)$ as a sum of atom-centred Gaussians,
\begin{equation}\label{eq:model}
  \hat{\rho}(\bvr) = \sum_{i,k} q_{ik}\, g_{ik}(\bvr),
\end{equation}
with normalised basis functions
\begin{equation}\label{eq:gaussian_real}
  g_{ik}(\bvr) = \frac{1}{(\sqrt{\pi}\,\sigma_k)^3}
  \exp\!\Bigl(-\frac{|\bvr - \vR_i|^2}{\sigma_k^2}\Bigr),
\end{equation}
where the index $i$ runs over atomic sites at positions~$\vR_i$
and $k = 1,\ldots,N_\text{GPA}$ (where GPA is gaussians per atom) labels Gaussians of different widths~$\sigma_k$ at each site.
In reciprocal space this model density is represented as,
\begin{equation}\label{eq:gG}
  g_{ik}(\vG) = \frac{1}{V}\,e^{-i\vG\cdot\vR_i}\,
  \exp\!\Bigl(-\frac{|\vG|^2\,\sigma_k^2}{4}\Bigr),
\end{equation}
where $V$ is the unit-cell volume, so that the reconstructed 
model charge density in reciprocal space is
\begin{equation}\label{eq:rho_model}
  \hat{\rho}(\vG) = \sum_{i,k} q_{ik}\, g_{ik}(\vG).
\end{equation}
where $q_{ik}$ is the charge corresponding to a Gaussian centred around atom $i$ with width $k$.
Conventionally, the Gaussian widths are arranged in a geometric series,
\begin{equation}\label{eq:sigma_series}
  \sigma_k = \sigma_{\text{start}}\cdot f^{\,k-1},
  \quad k=1,\ldots,N_\text{GPA},
\end{equation}
controlled by the starting width~$\sigma_\text{start}$ and
spacing factor~$f$ are obtained by minimising the following fitting function, $\mathcal{F}$, ~\cite{blochl1995}
\begin{align}\label{eq:functional}
  \mathcal{F}(\{q_{ik}\},\lambda)
  &= \frac{V}{2}\sum_{\vG\neq\mathbf{0}}
     w(\vG)\,
     \Bigl|\rho(\vG) - \sum_{ik}q_{ik}\,g_{ik}(\vG)\Bigr|^2
     \nonumber\\
  &\quad - \lambda\Bigl[V\rho(\mathbf{0})
     - \sum_{ik}q_{ik}\,V g_{ik}(\mathbf{0})\Bigr],
\end{align}
where $\lambda$ is a Lagrange multiplier enforcing charge
conservation and $w(\vG)$ is a weight function that biases the
fit toward the low-$\vG$ region based on a reciprocal space cutoff given by $g_c$.
Two technical challenges arise in applying this formalism: choosing the reciprocal-space cutoff~$g_c$ in $w(\vG)$, and ensuring numerical stability of the resulting linear system.
We address both of these challenges in subsequent sub-sections.


\subsection{Choosing the reciprocal-space cutoff~$g_c$}%
\label{sec:weight}

The weight function that biases the fit toward the multipole-carrying low-$\vG$ components is given by,
\begin{equation}\label{eq:weight}
  w(\vG) =
  \begin{cases}
    \displaystyle
    \frac{4\pi\,(|\vG|^2 - g_c^2)^2}{|\vG|^2\, g_c^2},
    & 0 < |\vG| < g_c, \\[6pt]
    0, & \text{otherwise}.
  \end{cases}
\end{equation}
This function diverges as $|\vG|\to 0$, ensuring that the monopole and low-order multipoles are reproduced with high fidelity, and decays smoothly to zero at the cutoff~$g_c$, beyond which reciprocal-space information is discarded.

\begin{figure}[!htb]
  \centering
  \includegraphics[width=\columnwidth]{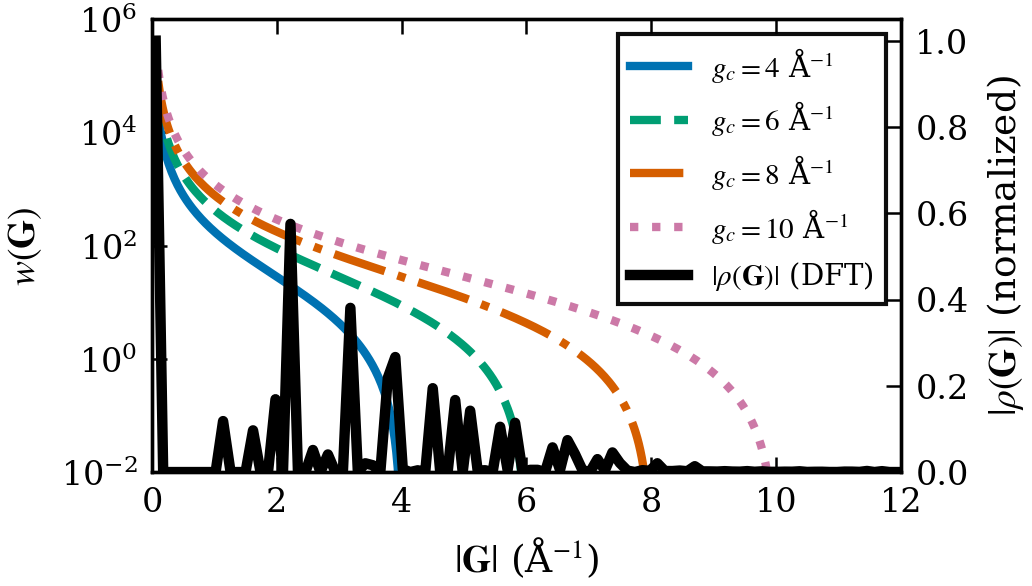}
  \caption{Weight function $w(\vG)$ [Eq.~\eqref{eq:weight}] for representative values of the reciprocal-space cutoff~$g_c$ (colored curves, left axis). The black curve shows the DFT charge density $|\rho(\vG)|$ in reciprocal space (right axis), with peaks at allowed reciprocal lattice vectors. The cutoff $g_c$ determines how many of these peaks are included in the fit.}
  \label{fig:weight}
\end{figure}
The physical motivation for this choice of function is clear from the Taylor expansion of $\rho(\vG)$ near the origin:
\begin{equation}\label{eq:taylor}
  \rho(\vG) = \rho(\mathbf{0}) + i\vG\cdot\mathbf{p}
            - \frac{1}{2}G_\alpha G_\beta Q_{\alpha\beta} + \cdots,
\end{equation}
where $\rho(\mathbf{0}) = N_e/V$ is the monopole (total charge), $\mathbf{p} = \int \bvr\,\rho(\bvr)\,d^3r$ is the dipole moment, and $Q_{\alpha\beta}$ is the quadrupole tensor.  These low-order moments fully determine the long-range electrostatic interaction between well-separated charge distributions.

The cutoff $g_c$ controls the fitting bandwidth.
Figure~\ref{fig:weight} illustrates this parameter dependence by overlaying the weight function with the DFT charge density in reciprocal
space.
The charge density $|\rho(\vG)|$ exhibits sharp peaks at discrete $|\vG|$ values corresponding to the allowed reciprocal lattice vectors, with spacing set by the lattice constant.  Larger $g_c$ (e.g., 10\,\AA$^{-1}$) includes more of these peaks but risks over-fitting high-frequency structure that atom-centered Gaussians cannot faithfully represent, leading to ill-conditioned linear systems.
Smaller $g_c$ (e.g., 4\,\AA$^{-1}$) improves numerical stability but may exclude important structural information.  


\subsection{Numerical instability of constrained optimisation}\label{sec:kkt}

The numerical issues associated with the DDAP method arise from the linear system obtained by setting the derivatives
$\partial\mathcal{F}/\partial q_{ik} = 0$ and
$\partial\mathcal{F}/\partial\lambda = 0$:
\begin{equation}\label{eq:kkt}
  \begin{pmatrix} \mA & \vc \\ \vc^T & 0 \end{pmatrix}
  \begin{pmatrix} \vq \\ \lambda \end{pmatrix}
  =
  \begin{pmatrix} \vb \\ N_e \end{pmatrix},
\end{equation}
where, using a composite index $\alpha\equiv(i,k)$,
\begin{subequations}\label{eq:Ab}
\begin{align}
  A_{\alpha\beta}
    &= V \sum_{\vG \neq \mathbf{0}}
       w(\vG)\, g_\alpha^*(\vG)\, g_\beta(\vG),
  \label{eq:Amat}\\
  b_\alpha
    &= V \sum_{\vG \neq \mathbf{0}}
       w(\vG)\, \rho^*(\vG)\, g_\alpha(\vG),
  \label{eq:bvec}\\
  c_\alpha
    &= V\, g_\alpha(\vG{=}\mathbf{0}) = 1,
  \label{eq:cvec}\\
  N_e
    &= V\,\rho(\vG{=}\mathbf{0}).
  \label{eq:Ne}
\end{align}
\end{subequations}
When $\mA$ is well-conditioned, Equation ~\eqref{eq:kkt} has the
closed-form solution~\cite{blochl1995}
\begin{equation}\label{eq:kkt_sol}
  q_\alpha = \sum_\beta (A^{-1})_{\alpha\beta}
  \biggl(b_\beta - c_\beta\,
  \frac{\sum_{kl}c_k(A^{-1})_{kl}b_l - N_e}%
       {\sum_{mn}c_m(A^{-1})_{mn}c_n}\biggr).
\end{equation}
As Equation \eqref{eq:kkt_sol} involves~$\mA^{-1}$, roundoff errors are amplified by the condition number~$\kappa(\mA)$. 
For over-complete bases (high $N_\text{GPA}$) or poorly chosen widths, $\kappa(\mA)$ can exceed $10^{15}$, at which point the charges diverge to $\mathcal{O}(10^{2})$\,electrons and become physically meaningless.  


\subsection{Numerical baseline of DDAP charge determination}\label{sec:trad}

In this section, we establish a baseline and illustrate the parameter sensitivity of the traditional DDAP algorithm.
As a basis for comparison, we use the reconstructed charge density, defined as
\begin{equation}
    \hat{\rho}(\mathbf{G}) = \sum_\alpha q_\alpha g_\alpha(\mathbf{G})
\end{equation} 

\begin{figure*}[!htb]
  \centering
  \includegraphics[width=\textwidth]{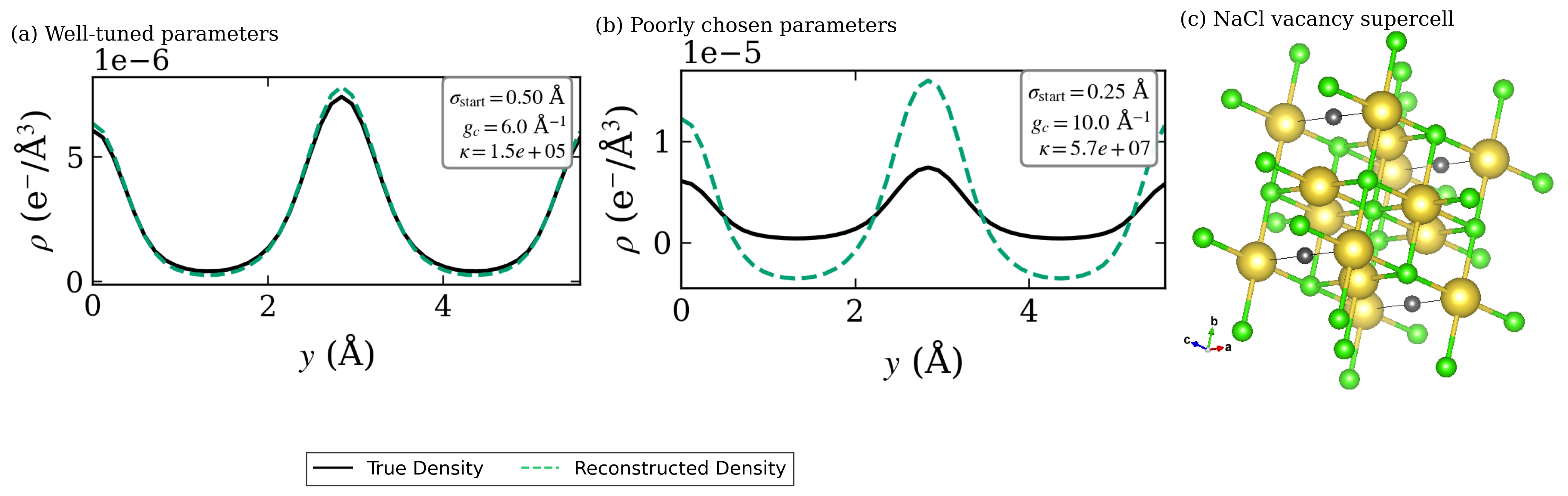}
  \caption{Sensitivity of traditional DDAP to parameter choice.
  (a)~Well-tuned parameters ($\sigma_\text{start}=0.50$\,\AA, $f=1.25$,
  $g_c=6.0$\,\AA$^{-1}$) yield excellent reconstruction with
  $\kappa \approx 10^{5}$.
  (b)~Poorly chosen parameters produce severe reconstruction errors with
  $\kappa \approx 10^{7}$.
  (c)~NaCl $1{\times}1{\times}1$ vacancy supercell used for validation;
  gold and green spheres denote Na and Cl, grey spheres mark vacant Cl sites.}
  \label{fig:trad_recon}
\end{figure*}

Figure ~\ref{fig:trad_recon} shows the true and reconstructed charge density of a defected-NaCl $1{\times}1{\times}1$ system.
With well-tuned parameters (Figure \ref{fig:trad_recon}(a)), the traditional solver achieves excellent agreement with the true
reference density.
The planar-averaged charge density (dashed green line) from DDAP closely tracks the the same quantity from DFT (solid black line) across the entire cell, with the characteristic peaks at atomic positions reproduced correctly. 
The condition number
$\kappa(\mA) \approx 1.5 \times 10^{5}$ remains well within the
numerically stable regime.

The reconstruction quality is sensitive to the choice of $(\sigma_\text{start},f,g_c)$.
Figure~\ref{fig:trad_recon}(b) demonstrates this fragility.
Reducing $\sigma_\text{start}$ from 0.50 to 0.25\,\AA\ and increasing $g_c$ from 6.0 to 10.0\,\AA$^{-1}$ leads to catastrophic failure to reconstruct the true charge density.
The DDAP density exhibits incorrect peak amplitudes, with the central peak underestimated
by nearly 50\% and the oscillation structure severely distorted.
The condition number increases to $\kappa \approx 5.7 \times 10^{7}$, indicating the onset of numerical ill-conditioning.

These results highlight a fundamental limitation of the traditional approach: no systematic procedure exists to determine the optimal parameter set for a given system.
Users must rely on heuristics or trial-and-error, with no guarantee that chosen values will transfer across different chemistries, supercell sizes, or defect configurations.
This lack of a systematic approach motivates \textit{opt}-DDAP a differentiable framework, which automates
parameter selection through gradient-based optimisation.

\section{\textit{opt}-DDAP creates numerically stable, parameter insensitive charges} \label{sec:autoddap_algo}
\subsection{Pseudoinverse followed by normalisation}\label{sec:pinv}

To address the numerical instability of the constrained solver, we replace the constrained system with an unconstrained least-squares solution obtained via the Moore-Penrose pseudoinverse,

\begin{equation}\label{eq:pinv}
  q_\alpha^{\,\text{raw}} = (\mA^{+}\,\vb)_\alpha,
\end{equation}

where $\mA^{+}$ is computed from the singular-value decomposition of~$\mA$ with a relative tolerance that discards near-zero singular values. 
This variable transformation yields the minimum-norm solution and remains numerically stable even  when $\kappa(\mA) \gg 10^{10}$.
Charge conservation is subsequently enforced by rescaling,
\begin{equation}\label{eq:renorm}
  q_\alpha = q_\alpha^{\,\text{raw}} \times
  \frac{N_e}{\sum_\beta q_\beta^{\text{raw}}}.
\end{equation}
\\
Numerical stability of $\mA$ is essential not only for physical 
correctness of the charges, but also for reliable automatic 
differentiation: ill-conditioned matrices produce unstable 
gradients in the backward pass through the pseudoinverse solve.
By discarding near-zero singular values, the pseudoinverse 
stabilizes both the forward solve and the gradient flow.

While the pseudoinverse resolves the numerical instability, it does not determine the optimal basis parameters.
In particular, the choice of~$g_c$ and
$\sigma_\text{start}$ remains ambiguous, motivating the optimisation strategy described next.

\subsection{Differentiable DDAP and parameter optimization}%
\label{sec:optim}

One of the central contributions of this work is to cast the entire DDAP pipeline, weight function evaluation, Gaussian basis construction, matrix assembly, pseudoinverse solve, and charge renormalization, as a differentiable computational graph in PyTorch~\cite{pytorch}.
The three continuous parameters
\begin{equation}\label{eq:params}
  \boldsymbol{\theta}
  = (\sigma_{\text{start}},\; f,\; g_c)
\end{equation}
control the Gaussian widths via Equation \eqref{eq:sigma_series} and
the reciprocal-space cutoff.
They are initialised to physically motivated defaults and optimised by minimising the unweighted
reconstruction loss
\begin{equation}\label{eq:loss}
  \mathcal{L}(\boldsymbol{\theta})
  = \sum_{\vG \neq \mathbf{0}}
    \bigl|\rho(\vG)
    - \hat{\rho}_{\boldsymbol{\theta}}(\vG)\bigr|^2,
\end{equation}
where
$\hat{\rho}_{\boldsymbol{\theta}}(\vG)
= \sum_\alpha q_\alpha(\boldsymbol{\theta})\,
  g_\alpha(\vG;\boldsymbol{\theta})$
is the model density reconstructed from the DDAP charges at the
current parameter values.
Note that $\mathcal{L}$ differs from the conventional approach (see Equation \eqref{eq:functional}) in that the Lagrange 
multiplier is replaced by the rescaling of 
Equation \eqref{eq:renorm}, making the loss an unconstrained 
scalar function of~$\boldsymbol{\theta}$.

Gradients $\nabla_{\boldsymbol{\theta}}\mathcal{L}$ are obtained by 
automatic differentiation through the pseudoinverse solve, and the 
parameters are updated with the Adam optimizer~\cite{adam}.
Box constraints restrict each parameter to a physically meaningful 
range, for example, $\sigma_{\text{start}} \in [0.2, 0.6]$\,\AA\ ensures the 
Gaussians are neither too narrow nor too 
wide; $f \in [1.15, 1.5]$ maintains 
adequate spacing between Gaussian widths; and 
$g_c \in [4.0, 12.0]$\,\AA$^{-1}$ keeps the fitting bandwidth 
within the physically relevant reciprocal-space range.
These bounds are enforced by clamping the parameters after each 
Adam step.
While the original DDAP approach employed a fixed spacing factor 
$f = 1.5$ for isolated molecules in vacuum supercells~\cite{blochl1995}, 
dense periodic solids have short nearest-neighbour distances where a large 
$\sigma_\text{max} = \sigma_\text{start}\cdot f^{N_\text{GPA}-1}$ 
causes significant overlap between basis functions on adjacent atoms, 
leading to ill-conditioned systems.
We therefore treat $f$ as an optimisable parameter, allowing the 
method to adapt automatically to the local atomic environment of 
each system (see SI)

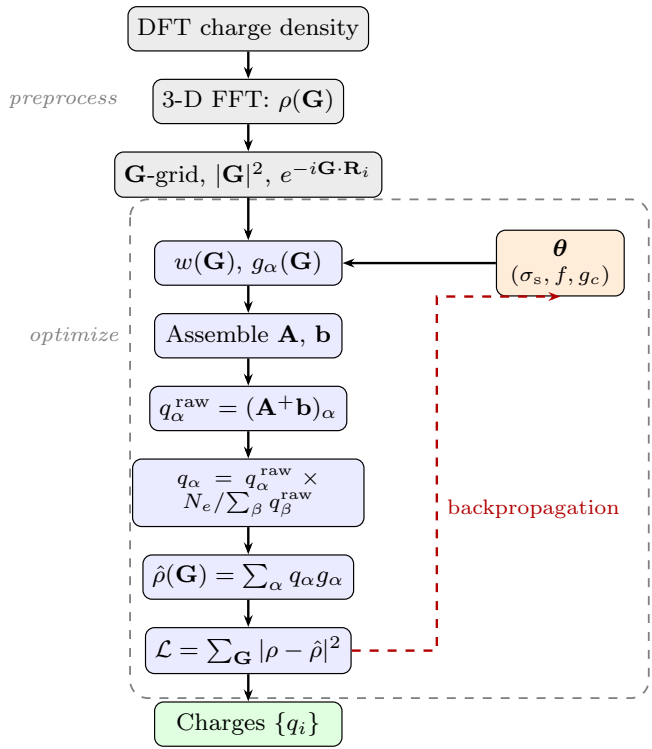
\begin{figure}[!htb]
\centering
\resizebox{\columnwidth}{!}{%
\begin{tikzpicture}[
  node distance=0.32cm and 0.5cm,
  box/.style={rectangle, draw, rounded corners=3pt,
              minimum width=2.2cm, minimum height=0.52cm,
              font=\footnotesize, align=center,
              fill=blue!8},
  prebox/.style={rectangle, draw, rounded corners=3pt,
              minimum width=2.2cm, minimum height=0.52cm,
              font=\footnotesize, align=center,
              fill=gray!15},
  param/.style={rectangle, draw, rounded corners=3pt,
                minimum width=1.5cm, minimum height=0.48cm,
                font=\footnotesize, align=center,
                fill=orange!15},
  output/.style={rectangle, draw, rounded corners=3pt,
                minimum width=2.2cm, minimum height=0.52cm,
                font=\footnotesize, align=center,
                fill=green!12},
  arr/.style={-{Stealth[length=4pt]}, thick},
  autograd/.style={-{Stealth[length=4pt]}, thick, dashed, red!70!black},
  label/.style={font=\scriptsize\itshape, text=gray},
]

\node[prebox] (chgcar) {DFT charge density};
\node[prebox, below=of chgcar] (fft) {3-D FFT: $\rho(\vG)$};
\node[prebox, below=of fft] (gvec) {$\vG$-grid, $|\vG|^2$, $e^{-i\vG \cdot \vR_i}$};

\node[label, left=0.3cm of fft] {preprocess};

\node[box, below=0.5cm of gvec] (weight) {$w(\vG)$, $g_\alpha(\vG)$};
\node[box, below=of weight] (assemble) {Assemble $\mA$, $\vb$};
\node[box, below=of assemble] (solve)
  {$q_\alpha^{\,\text{raw}} = (\mA^+\vb)_\alpha$};
\node[box, below=of solve, font=\scriptsize, minimum height=0.72cm,
      text width=2.5cm, align=center] (renorm)
  {$q_\alpha = q_\alpha^{\,\text{raw}} \times N_e/{\textstyle\sum_\beta q_\beta^{\text{raw}}}$};
\node[box, below=of renorm] (recon) {$\hat\rho(\vG) = \sum_\alpha q_\alpha g_\alpha$};
\node[box, below=of recon] (loss) {$\mathcal{L} = \sum_{\vG}|\rho - \hat\rho|^2$};

\node[label, left=0.3cm of assemble] {optimize};

\node[output, below=of loss] (charges) {Charges $\{q_i\}$};

\node[param, right=1.8cm of weight] (theta) {$\boldsymbol{\theta}$\\[-1pt]{\scriptsize$(\sigma_\text{s},f,g_c)$}};

\draw[arr] (chgcar) -- (fft);
\draw[arr] (fft) -- (gvec);

\draw[arr] (gvec) -- (weight);
\draw[arr] (weight) -- (assemble);
\draw[arr] (assemble) -- (solve);
\draw[arr] (solve) -- (renorm);
\draw[arr] (renorm) -- (recon);
\draw[arr] (recon) -- (loss);
\draw[arr] (loss) -- (charges);

\draw[arr] (theta.west) -- (weight.east);

\draw[autograd]
  (loss.east) -- ++(1.0,0) 
  |- (theta.south)
  node[pos=0.20, right, font=\scriptsize, text=red!70!black] {backpropagation};

\begin{scope}[on background layer]
  \draw[
    dashed, gray, rounded corners=6pt, line width=0.6pt
  ]
  ([xshift=-8pt, yshift=38pt] assemble.north west)
  rectangle
  ([xshift=8pt, yshift=-8pt] loss.south -| theta.east);
\end{scope}

\end{tikzpicture}%
}
\caption{Data flow of the differentiable DDAP pipeline.
Gray boxes denote preprocessing steps executed once;
blue boxes form the optimization loop.
The parameter block~$\boldsymbol{\theta} = (\sigma_\text{start}, f, g_c)$
controls both the Gaussian basis and the weight function cutoff.
The dashed red arrow indicates gradient back-propagation via
automatic differentiation.}
\label{fig:flowchart}
\end{figure}

Figure \ref{fig:flowchart} shows the steps required to determine the \textit{opt}-DDAP charges. 
The preprocessing steps (gray boxes), FFT of the DFT charge 
density and construction of the $\vG$-grid, are performed once.
The optimisation loop (blue boxes) then iterates the following loop: given the current parameters $\boldsymbol{\theta} = (\sigma_\text{start}, f, g_c)$, the weight function and Gaussian basis are evaluated, the matrix $\mA$ and vector $\vb$ are assembled, the pseudoinverse solver yields raw charges which are renormalised, and the reconstructed density $\hat{\rho}(\vG)$ is used to evaluate the loss $\mathcal{L}$.
Gradients are back-propagated through the entire pipeline (dashed  red arrow) to update $\boldsymbol{\theta}$ via the Adam optimiser~\cite{adam}.

\section{Validation tests} \label{sec:validation}
\subsection{Optimised DDAP Reconstruction}\label{sec:ionic}

\begin{figure*}[!htb]
  \centering
  \includegraphics[width=2.\columnwidth]{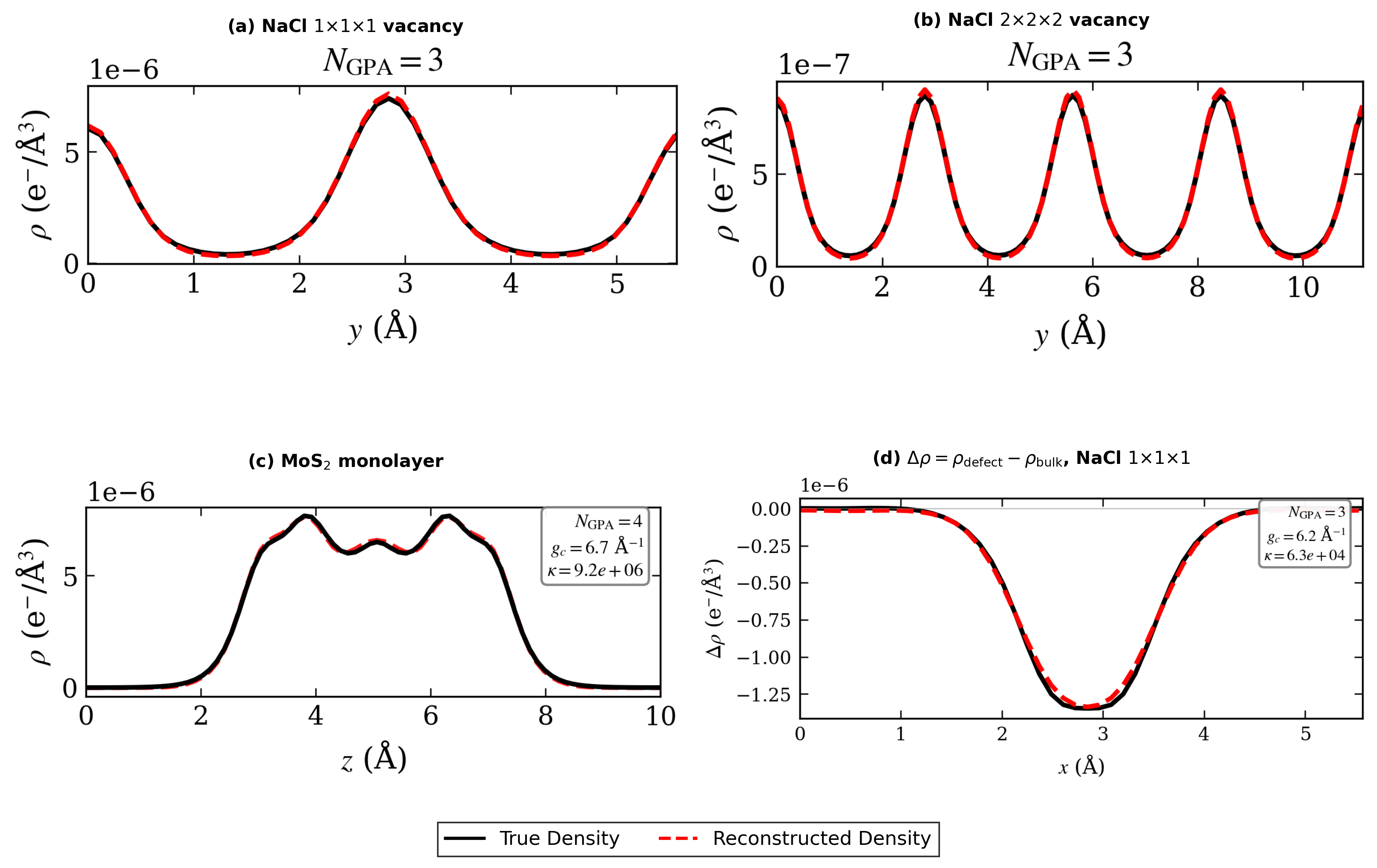}
  \caption{Optimized DDAP reconstructions.
  (a)~NaCl $1{\times}1{\times}1$ vacancy ($y$-profile).
  (b)~NaCl $2{\times}2{\times}2$ vacancy ($y$-profile).
  (c)~MoS$_2$ monolayer ($z$-profile).
  (d)~Difference density $\Delta\rho = \rho_\text{defect} - \rho_\text{bulk}$
  for NaCl $1{\times}1{\times}1$ ($x$-profile through the vacancy site).
  Black solid: VASP reference; red dashed: DDAP reconstruction.
  All calculations use $N_\text{GPA} = 3$.}
  \label{fig:opt_recon}
\end{figure*}

Figure~\ref{fig:opt_recon} presents the optimised DDAP
reconstructions for all test systems.  
Two chemically distinct systems were chosen: NaCl vacancy 
supercells in two sizes ($1{\times}1{\times}1$ and 
$2{\times}2{\times}2$) as a prototypical ionic system with 
well-localised atom-centred charges, and a MoS$_2$ monolayer 
as a covalent benchmark where charge extends into the 
interatomic bonding regions.
Additionally, the charge difference density 
$\Delta\rho = \rho_\text{defect} - \rho_\text{bulk}$ of the 
NaCl $1{\times}1{\times}1$ vacancy is included to test the 
sensitivity of the framework to small, spatially localised 
charge redistributions induced by a point defect.
In each panel, the planar-averaged charge density from DDAP (red dashed) is compared against the DFT reference (black solid, see Section \ref{sec:compdetails} for more details on the computational methods).

For the ionic NaCl vacancy systems (Figure \ref{fig:opt_recon}a,b),
the agreement is excellent across the entire spatial range.  The
$1{\times}1{\times}1$ supercell (7 atoms) and the larger
$2{\times}2{\times}2$ supercell (63 atoms) both show near-perfect
overlap between the DDAP and true densities, with characteristic peaks at atomic positions faithfully reproduced. 
The optimizer converges to similar parameters for both supercell sizes: $\sigma_\text{start} \approx 0.50$\,\AA\ and
$g_c \approx 6$\,\AA$^{-1}$, demonstrating transferability across system sizes.
Condition numbers remain well within the stable regime ($\kappa \sim 10^{3}$--$10^{5}$).

The MoS$_2$ monolayer (Figure \ref{fig:opt_recon}c) presents a more challenging test due to its covalent bonding character.
Unlike ionic systems where charge is localised on atomic sites,
MoS$_2$ exhibits significant bonding charge in the interatomic
regions.  The $z$-profile shows the characteristic double-peak
structure corresponding to the S--Mo--S trilayer, with the DDAP
reconstruction capturing the overall envelope despite the
inherent limitation of atom-centered basis functions.

The most stringent test of physical fidelity is the
reconstruction of the difference density
$\Delta\rho(\bvr) = \rho_\text{defect}(\bvr) - \rho_\text{bulk}(\bvr)$,
which isolates the charge redistribution induced by the Cl
vacancy (Figure \ref{fig:opt_recon}d). 
This quantity is considerably smaller in magnitude than the total density and exhibits both positive and negative regions, making it highly sensitive to fitting errors.
To compute the DDAP charges, we apply the algorithm independently to $\rho_\text{defect}$ and $\rho_\text{bulk}$ using \emph{bulk} atomic positions in both cases, ensuring that the difference reflects only electronic
redistribution.

The agreement in Figure \ref{fig:opt_recon}d is excellent.
The DDAP curve closely tracks the DFT reference, capturing the
sharp negative feature centred on the vacancy site
($x \approx 2.8$\,\AA) that corresponds to the missing Cl
valence electrons. 
The condition number for the difference density ($\kappa \approx 1.1 \times 10^{5}$) remains acceptable, and the integrated charge difference matches the expected value
of $-7$\,e$^-$ (one missing Cl$^-$ ion).
This validates that the optimized DDAP framework preserves physically meaningful charge redistribution associated
with point defects, a capability essential for machine-learning
potentials that must capture defect-induced charge transfer.

\subsection{Robustness to initial conditions}\label{sec:robust}

A key challenge for any gradient-based optimisation is whether the algorithm converges to the same solution from different starting points, or whether multiple local minima exist in the loss landscape.
To address this challenge, we initialise the optimizer from four distinct parameter combinations spanning the
physically reasonable range:
$(\sigma_0, g_c) \in \{(0.4, 11), (0.5, 8), (0.6, 6), (0.7, 5)\}$
(units: \AA\ and \AA$^{-1}$, respectively), applied to the 
NaCl $1{\times}1{\times}1$ vacancy system.

\begin{figure}[!htb]
  \centering
  \includegraphics[width=\columnwidth]{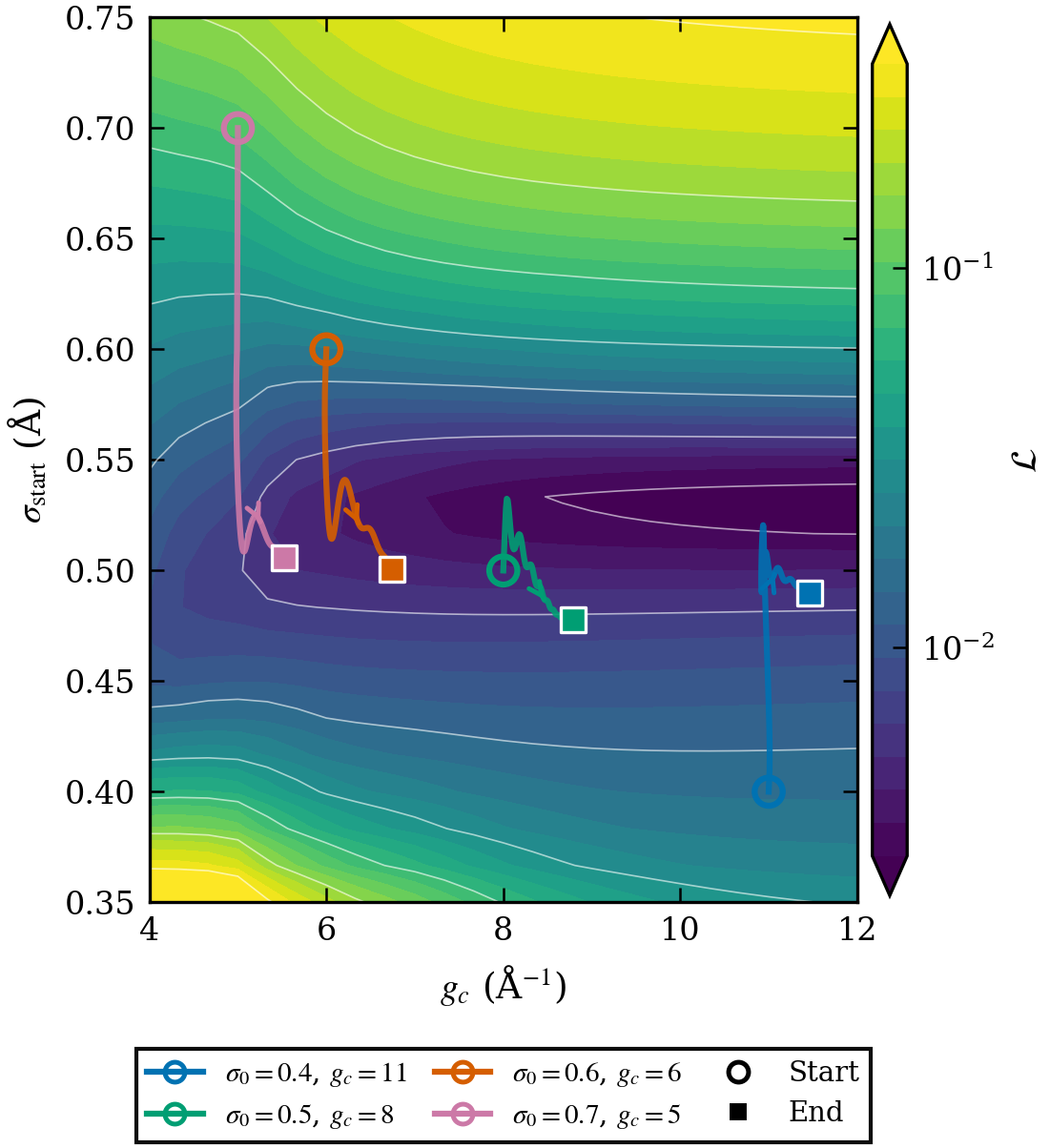}
  \caption{Optimisation trajectories from four different initial
  parameter sets for NaCl $1{\times}1{\times}1$, overlaid on the
  loss landscape $\mathcal{L}(\sigma_\text{start}, g_c)$.  Open
  circles mark starting points; filled squares mark converged
  solutions.  All trajectories converge to the same basin near
  $(\sigma_\text{start}, g_c) \approx (0.50\,\text{\AA},
  7\text{--}11\,\text{\AA}^{-1})$.}
  \label{fig:starting_pts}
\end{figure}

Figure~\ref{fig:starting_pts} shows the resulting optimisation
trajectories overlaid on the loss landscape
$\mathcal{L}(\sigma_\text{start}, g_c)$ after 100 Adam iterations
with learning rate $\eta = 0.01$.  The loss landscape exhibits a
broad minimum region (dark purple) where reconstruction quality
is optimal.  All four trajectories converge toward
$\sigma_\text{start} \approx 0.50$\,\AA, though the final $g_c$
values show some variation depending on the starting point.

\begin{table}[!htb]
\caption{Convergence from different initial conditions for NaCl
$1{\times}1{\times}1$ ($N_\text{GPA} = 3$). Despite widely
varying starting points, all runs converge to similar
$\sigma_\text{start}^*$ values and yield consistent atomic
charges. $q_\text{Cl}$ denotes the charge Cl site.}
\label{tab:robustness}
\begin{ruledtabular}
\begin{tabular}{cccccc}
\multicolumn{2}{c}{Initial} & \multicolumn{2}{c}{Final} & & \\
\cline{1-2} \cline{3-4}
$\sigma_0$ & $g_c^{(0)}$ & $\sigma^*$ & $g_c^*$ & $q_\text{Cl}$\\
(\AA) & (\AA$^{-1}$) & (\AA) & (\AA$^{-1}$) & (e$^-$) \\
\hline
0.40 & 11.0 & 0.490 & 11.5 & 7.458 \\
0.50 & 8.0  & 0.478 & 8.8  & 7.543 \\
0.60 & 6.0  & 0.500 & 6.7  & 7.509 \\
0.70 & 5.0  & 0.506 & 5.5  & 7.574 \\
\end{tabular}
\end{ruledtabular}
\end{table}
Table~\ref{tab:robustness} summarizes the convergence results. We find three key results corresponding to the behaviour of the optimiser.
First, all runs converge to
$\sigma_\text{start}^* \in [0.478, 0.505]$\,\AA, a spread of
only 0.027\,\AA\ (5.4\%), indicating that this parameter is
well-determined by the optimization.  Second, the final $g_c^*$
values show greater variation, ranging from 5.5 to
11.5\,\AA$^{-1}$, suggesting that the loss landscape is
relatively flat along this direction once $\sigma_\text{start}$
is near its optimal value.  Third, despite these differences in
$g_c^*$, the resulting Cl charges $q_\text{Cl}$ are remarkably consistent at $7.52 \pm 0.05$\,e$^-$, and Na charges
(not shown) similarly converge to $6.61 \pm 0.04$\,e$^-$.

The lowest loss ($\mathcal{L}^* = 1.8 \times 10^{-3}$) is
achieved from the $(0.4, 11)$ starting point, while the highest
($\mathcal{L}^* = 4.3 \times 10^{-3}$) comes from $(0.7, 5)$.
This 2.4-fold difference in loss translates to only a 1.6\%
variation in the extracted charges, demonstrating that the
physical observables are robust even when the optimizer settles
into slightly different regions of the loss landscape.  This
robustness is essential for practical applications where good
initial guesses may not be available.
These results suggest that $\sigma_\text{start}$ is the 
physically dominant parameter. Once it converges to its optimal 
value near $0.50$\,\AA, the reconstruction quality is largely 
insensitive to the precise value of $g_c$, which is consistent 
with Bl\"ochl's original observation that results are virtually 
insensitive to the cutoff within a reasonable range~\cite{blochl1995}.

The robustness of \textit{opt}-DDAP to initial conditions makes 
it well-suited for high-throughput workflows.
The algorithm can be applied systematically to large databases 
of DFT calculations, such as those generated by the 
Materials Project~\cite{materialsproject2013} or 
AFLOW~\cite{curtarolo2012aflow}, to produce physically 
grounded atomic charges at scale, without manual parameter 
tuning for each system.
These charges can directly serve as inputs to MLIPs that 
incorporate long-range electrostatics via Ewald 
summation~\cite{ewald1921,cheng2025les}, addressing a key bottleneck 
in developing MLIPs for ionic and partially ionic materials.
Future extensions will explore incorporating the DDAP loss 
directly into the MLIP training objective, enabling 
simultaneous optimisation of short-range and long-range 
interactions.

\section{Conclusion}\label{sec:conclusion}

In this work, we describe \textit{opt}-DDAP, a differentiable reformulation of the density-derived atomic point charge (DDAP) method that enables automatic optimisation of the Gaussian basis parameters $(\sigma_\text{start}, f, g_c)$. The key changes to the conventional DDAP method are: (1) we replace the numerically fragile constrained solver with a pseudo-inverse solution followed by charge renormalization, which remains stable even for condition numbers $\kappa(\mA) > 10^{10}$ and (2) we implement the entire pipeline as a differentiable computational graph in PyTorch, allowing end-to-end optimisation via automatic differentiation.

We validate \textit{opt}-DDAP on NaCl vacancy supercells and monolayer MoS$_2$. We show that the optimised charges faithfully reproduce the low-$\vG$ multipole structure of the DFT charge density.
The optimizer converges robustly from diverse initial conditions to a consistent solution, eliminating the need for system-specific parameter tuning.
Crucially, the framework accurately captures the difference density, validating its applicability to defected systems where charge redistribution is highly localised.

\FloatBarrier

\section{Computational details}\label{sec:compdetails}

All DFT calculations were performed with the Vienna \textit{ab initio}
Simulation Package (VASP)~\cite{vasp}.
Three test systems were considered in this work:
(i)~a $1{\times}1{\times}1$ NaCl rocksalt supercell with a
single Cl vacancy (7~atoms),
(ii)~a $2{\times}2{\times}2$ NaCl supercell with a single Cl
vacancy (63~atoms) and
(iii)~a monolayer MoS$_2$ unit cell.

The structure, cell and volume of bulk NaCl and an MoS$_2$ were relaxed using the \verb|ISIF=3| option implemented in VASP using the PBE functional with a \textit{k}-point sampling of 12$\times$12$\times$12, yielding a lattice constant of $a=3.93~\mathrm{\AA}$, $c=5.56~\mathrm{\AA}$ for NaCl and $a=3.18$ for MoS$_2$ with 12$\r{AA}$ along the vacuum dimension.
The \verb|Na_pv|, \verb|Cl|, \verb|Mo|, \verb|S| PAW potential were used for Na, Cl, Mo and S atoms, respectively.
A plane-wave cutoff of 262~eV was used for all charged defect calculations.
The \textit{k}-point density was kept constant for different defect concentrations on NaCl by dividing the number of repetitions of the cell by 12. Gaussian smearing was used with a broadening on 0.1~eV for DFT calculations.

For the vacancy systems, bulk reference calculations on the
corresponding pristine supercell were also performed to
construct the difference density
$\Delta\rho = \rho_\text{defect} - \rho_\text{bulk}$.

The DDAP optimization used $N_\text{GPA} = 3$, with initial
$\sigma_\text{start} = 0.5$~\AA, $f = 1.25$, and
$g_c = 6.0$~\AA$^{-1}$.  Adam was run with learning rate
$\eta = 10^{-2}$ for up to 100~epochs.  Box constraints were
$\sigma_\text{start}\in[0.2,\,0.6]$~\AA,
$f\in[1.15,\,1.5]$, and
$g_c\in[4.0,\,12.0]$~\AA$^{-1}$.  All tensors used 64-bit
floating-point arithmetic (\texttt{torch.float64}).

\section{Code availability}
The \textit{auto}-DDAP implementation is available at \url{https://gitlab.com/electrocatalysis-group/atomic-recipes}

\begin{acknowledgments}
S.V. acknowledges funding from ANRF MATRICS (ANRF/ARGM/2025/000139/TS) and seed grant (RD/0524-IRCCSH0-023) at IIT Bombay. All authors acknowledge compute resources from NSM PARAM Rudra HPC facility at IIT Bombay, which is implemented by C-DAC and supported by the Ministry of Electronics and Information Technology (MeitY) and Department of Science and Technology (DST), Government of India. The authors thank Prof.\ Ananth Govind Rajan (IISc Bangalore) for helpful discussions.
During the preparation of this work, the authors used Claude to help with paper writing and to refactor text for readability. After using this language model, the authors reviewed and edited the content as needed.
\end{acknowledgments}

\bibliographystyle{apsrev4-1}
\bibliography{references}

@article{Teale2022,
  author  = {Teale, Andrew M. and De Proft, Frank and Geerlings, Paul and Helgaker, Trygve and others},
  title   = {DFT exchange: sharing perspectives on the workhorse of quantum chemistry and materials science},
  journal = {Physical Chemistry Chemical Physics},
  year    = {2022},
  volume  = {24},
  issue   = {32},
  pages   = {18441-18548},
  publisher = {The Royal Society of Chemistry},
  doi     = {10.1039/D2CP02827A},
  url     = {https://doi.org}
}

@article{mueller2016high,
  author  = {Mueller, Tim and Hautier, Geoffroy and Jain, Anubhav and Ceder, Gerbrand},
  title   = {High-throughput computational design of cathode coatings for {Li}-ion batteries},
  journal = {Chemistry of Materials},
  volume  = {28},
  number  = {24},
  pages   = {8926--8934},
  year    = {2016},
  doi     = {10.1021/acs.chemmater.6b03361}
}

@article{aykol2016network,
  author  = {Aykol, Muratahan and Kim, Sunghyun and Hegde, Vinay I. and Snydacker, David and others},
  title   = {Network-based strategies for identifying electrode materials for multivalent batteries},
  journal = {Nature Communications},
  volume  = {7},
  number  = {1},
  pages   = {13779},
  year    = {2016},
  doi     = {10.1038/ncomms13779}
}

@article{winther2019catalysis,
  author  = {Winther, Kirsten T. and Hoffmann, Max J. and Boes, Jacob R. and Mamun, Osman and others},
  title   = {Catalysis-{H}ub.org, an open electronic structure database for surface reactions},
  journal = {Scientific Data},
  volume  = {6},
  number  = {1},
  pages   = {75},
  year    = {2019},
  doi     = {10.1038/s41597-019-0081-y}
}

@article{kitchin2018machine,
  author  = {Kitchin, John R.},
  title   = {Machine learning in catalysis},
  journal = {Nature Catalysis},
  volume  = {1},
  number  = {4},
  pages   = {230--232},
  year    = {2018},
  doi     = {10.1038/s41929-018-0056-y}
}

@article{materialsproject2013,
  author  = {Jain, Anubhav and Ong, Shyue Ping and Hautier, Geoffroy
             and Chen, Wei and Richards, William Davidson and Dacek, Stephen
             and Cholia, Shreyas and Gunter, Dan and Skinner, David
             and Ceder, Gerbrand and Persson, Kristin A.},
  title   = {Commentary: {The Materials Project}: A materials genome approach
             to accelerating materials innovation},
  journal = {APL Mater.},
  volume  = {1},
  pages   = {011002},
  year    = {2013},
  doi     = {10.1063/1.4812323}
}

@article{curtarolo2012aflow,
  author  = {Curtarolo, Stefano and Setyawan, Wahyu and Hart, Gus L.~W.
             and Jahnatek, Michal and Chepulskii, Roman V. and Taylor, Richard H.
             and Wang, Shidong and Xue, Junkai and Yang, Kesong and Levy, Ohad
             and Mehl, Michael J. and Stokes, Harold T.
             and Demchenko, Denis O. and Morgan, Dane},
  title   = {{AFLOW}: An automatic framework for high-throughput materials discovery},
  journal = {Comput.\ Mater.\ Sci.},
  volume  = {58},
  pages   = {218--226},
  year    = {2012},
  doi     = {10.1016/j.commatsci.2012.02.005}
}

@book{bader1990,
  author    = {Bader, Richard F. W.},
  title     = {Atoms in Molecules: A Quantum Theory},
  publisher = {Clarendon Press},
  address   = {Oxford},
  year      = {1990}
}

@article{mulliken1955,
  author  = {Mulliken, Robert S.},
  title   = {Electronic Population Analysis on {LCAO--MO} Molecular Wave
             Functions. {I}},
  journal = {J.\ Chem.\ Phys.},
  volume  = {23},
  pages   = {1833--1840},
  year    = {1955},
  doi     = {10.1063/1.1740588}
}

@article{glendening2012,
  author  = {Glendening, Eric D. and Landis, Clark R. and Weinhold, Frank},
  title   = {Natural bond orbital methods},
  journal = {WIREs Comput.\ Mol.\ Sci.},
  volume  = {2},
  pages   = {1--42},
  year    = {2012},
  doi     = {10.1002/wcms.51}
}

@article{blochl1995,
  author  = {Bl\"ochl, P. E.},
  title   = {Electrostatic decoupling of periodic images of
             plane-wave-expanded densities and derived atomic point charges},
  journal = {J.\ Chem.\ Phys.},
  volume  = {103},
  pages   = {7422},
  year    = {1995}
}

@article{vasp,
  author  = {Kresse, G. and Furthm\"uller, J.},
  title   = {Efficient iterative schemes for ab initio total-energy
             calculations using a plane-wave basis set},
  journal = {Phys.\ Rev.\ B},
  volume  = {54},
  pages   = {11169},
  year    = {1996}
}

@article{pytorch,
  author  = {Paszke, A. and others},
  title   = {PyTorch: An imperative style, high-performance deep learning library},
  journal = {Adv.\ Neural Inf.\ Process.\ Syst.},
  volume  = {32},
  pages   = {8024},
  year    = {2019}
}

@article{adam,
  author  = {Kingma, D. P. and Ba, J.},
  title   = {Adam: A method for stochastic optimization},
  journal = {arXiv:1412.6980},
  year    = {2014}
}

@article{kocer2022,
  author  = {Ko\c{c}er, E. and Ko, T. W. and Behler, J.},
  title   = {Neural network potentials: A concise overview of methods,
             applications, and challenges},
  journal = {Annu.\ Rev.\ Phys.\ Chem.},
  volume  = {73},
  pages   = {163},
  year    = {2022}
}

@article{schmidt2010ionic,
  author  = {Schmidt, Jochen and Krekeler, Christian and Dommert, 
             Florian and Zhao, Yuanyuan and Berger, Robert and 
             Delle Site, Luigi and Holm, Christian},
  title   = {Ionic Charge Reduction and Atomic Partial Charges 
             from First-Principles Calculations of 
             1,3-Dimethylimidazolium Chloride},
  journal = {J.\ Phys.\ Chem.\ B},
  volume  = {114},
  pages   = {6150--6155},
  year    = {2010},
  doi     = {10.1021/jp910771q}
}

@article{kumar2024bnnt,
  author  = {Kumar, Shiv and Govind Rajan, Ananth},
  title   = {Predicting Quantum-Mechanical Partial Charges in 
             Arbitrarily Long Boron Nitride Nanotubes to Accurately 
             Simulate Nanoscale Water Transport},
  journal = {J.\ Chem.\ Theory Comput.},
  volume  = {20},
  pages   = {3298--3307},
  year    = {2024},
  doi     = {10.1021/acs.jctc.4c00080}
}

@article{govindrajan2018hbn,
  author  = {Govind Rajan, Ananth and Strano, Michael S. 
             and Blankschtein, Daniel},
  title   = {Ab Initio Molecular Dynamics and Lattice Dynamics 
             Based Force Field for Modeling Hexagonal Boron Nitride 
             in Mechanical and Interfacial Applications},
  journal = {J.\ Phys.\ Chem.\ Lett.},
  volume  = {9},
  pages   = {1780--1786},
  year    = {2018},
  doi     = {10.1021/acs.jpclett.7b03443}
}

@article{cheng2025les,
  author  = {Cheng, Bingqing and Kim, Dongjin and others},
  title   = {Latent Ewald summation for machine learning of long-range interactions},
  journal = {npj Computational Materials},
  year    = {2025},
  doi     = {10.1038/s41524-025-01577-7}
}

@article{ewald1921,
  author  = {Ewald, Paul Peter},
  title   = {Die Berechnung optischer und elektrostatischer Gitterpotentiale},
  journal = {Annalen der Physik},
  volume  = {369},
  number  = {3},
  pages   = {253--287},
  year    = {1921},
  doi     = {10.1002/andp.19213690304}
}

@article{dommert2012charge,
  author  = {Dommert, Florian and Wendler, Katharina and Berger, Robert
             and Delle Site, Luigi and Holm, Christian},
  title   = {Force-Field Development for Ionic Liquids Based on Quantum
             Chemical Calculations: 1-Ethyl-3-Methylimidazolium Chloride
             as a First Example},
  journal = {ChemPhysChem},
  volume  = {13},
  pages   = {1625--1637},
  year    = {2012},
  doi     = {10.1002/cphc.201100781}
}
\end{document}


\title{Supplemental Material: \textit{opt}-DDAP}
\maketitle

\section{S1.~Why the Gaussian spacing factor $f$ is treated as optimisable}
\label{sec:si_f}

A fixed spacing factor of $f = 1.5$ was used throughout in the original DDAP formulation.
This assumption was validated on isolated molecules in large vacuum supercells where
adjacent atomic centres are well-separated~\cite{blochl1995}.
In that limit, even the widest Gaussian
$\sigma_\text{max} = \sigma_\text{start} \cdot f^{N_\text{GPA}-1}$
contributes negligible electron density at neighbouring sites.

Dense periodic solids violate this assumption.
For the NaCl system, the nearest-neighbour Na--Cl distance is
$d \approx 2.77$\,\AA.
With $N_\text{GPA} = 3$ and $\sigma_\text{start} = 0.5$\,\AA,
Bl\"{o}chl's default gives
$\sigma_\text{max} = 0.5 \times 1.5^2 = 1.125$\,\AA,
which is ${\sim}40\%$ of the bond length.
At this width, Gaussians on adjacent sites overlap significantly,
introducing near-linear dependence among the columns of the overlap
matrix $\mA$ and causing its condition number to grow by several
orders of magnitude relative to a well-spaced basis.
The resulting charges become numerically unreliable even before
ill-conditioning formally sets in.

Treating $f$ as a differentiable, optimisable parameter allows the
algorithm to automatically select a spacing that keeps
$\sigma_\text{max} \ll d$, maintaining a well-conditioned $\mA$
without any system-specific manual tuning.

\bibliographystyle{apsrev4-1}
\bibliography{references}